\documentclass[aps,prl,reprint,superscriptaddress]{revtex4-1} 
\usepackage{graphicx}
\usepackage{longtable}
\usepackage{color}

\def\lsim{\, \rlap{$<$}{\lower 1.1ex\hbox{$\sim$}}\,}
\begin{document}

\preprint{YNU-COS-2015001}

\title{Thermally Producing and Weakly Freezing-out Dark Matter in Bouncing Universe}

\author{Changhong~Li}
\email[]{changhongli@ynu.edu.cn}
\affiliation{Department of Astronomy, Yunnan University,\\ Key Laboratory of Astroparticle Physics of Yunnan Province, \\No.2 Cuihu North Road, Kunming, China 650091}
\affiliation{Department of Physics, Nanjing University, \\
22 Hankou Road, Nanjing, China 210093}

\begin{abstract}
We investigate the production and freeze-out of dark matter with a constant thermally averaged cross-section in a generic bouncing universe framework. Our result shows that, there is a novel avenue that dark matter is produced thermally and take a weakly freezing-out process, besides two previously known cases, the {\it thermally production \& strongly freezing-out} case and the {\it non-thermally production \& weakly freezing-out} case, in which the relic abundance of dark matter are inverse and proportional to its cross-section respectively.  We calculated the relic abundance of dark matter for this new case, and find its  relic abundance is independent of its cross-section. We also present the cosmological constraints on the cross-section and mass relation of dark matter for this new case.   
\end{abstract}


\maketitle

Inflation has almost been crowned as it solves the {\it horizon problem} and {\it flatness problem}~\cite{Guth:1980zm}, generates nearly scale-invariant curvature spectrum~\cite{Mukhanov:1990me}, which agrees well with the current array of observations~\cite{Komatsu:2010fb, Ade:2013kta}. However, it suffers {\it initial singularity problem} and {\it fine-tuning problem}~\cite{Liddle:2000cg}, which renders it unreliable on which it stands~\cite{Borde:1993xh}. To address the Initial Singularity Problem which inflation scenario inevitably suffers, a concordance of effort have made in recent years   by utilizing a generic bouncing feature of early universe models inspired by underlying physics such as string phenomenology and quantum loop theory~~\cite{Wands:1998yp, Khoury:2001wf, Gasperini:2002bn, Creminelli:2006xe, Cai:2007qw, Cai:2008qw,  Wands:2008tv, Bhattacharya:2013ut}. And a  stable and scale-invariant curvature spectrum, compatible with the current observation, in the bouncing universe scenario rid of Initial Singularity and Fine-tuning Problems is recently obtained~\cite{Li:2011nj, Li:2013bha}. Therefore, we are well motived to turn our attention to the bouncing universe scenario(See~\cite{Novello:2008ra, Brandenberger:2012zb} for recent reviews.) .

However, the preciously measured CMB spectra as well as scalar-tensor ratio may not be enough for distinguishing the inflationary and bouncing universe concretely, due to two respects: 1) the well-establised duality between inflationary and bouncing universe empowers both of them to generate the stable scale-invariant curvature power spectrum, which is compatible with current observation, with the same probability in the unified parameter space~\cite{Wands:1998yp, Finelli:2001sr, Boyle:2004gv,Li:2012vi, Li:2013bha}, therefore, the CMB spectra may not be able to serve as the direct evidence for neither inflationary nor bouncing universe; and 2) so far all of the inflationary and bouncing universe models are utilizing some fields undetected yet to drive inflation or big bounce at the early stage of the cosmological evolution. Hence their predictions about CMB spectra and the scalar-tensor ratio, which are built upon the linear perturbation theory of these unconfirmed fields, is still questionable. Therefore, we are motivated furtherly to investigate a concrete way for distinguishing the inflationary and bouncing universe with the recent and near-furture experimentally detectable evidences. 

Hopefully, the abundance of each subatomic particle and light chemical element would serve as a good candidate for investigating the early history of our universe---enjoying the successful philosophy and experience of Big Bang Theory. However, for these subatomic particles and light chemical elements, their synthesis and thermal decoupling happened below $10~$MeV, the energy scale of Big Bang Nucleosynthesis, which is much lower than the typical energy scale of inflation and big bounce~\cite{Dodelson:2003ft, Mukhanov:2005sc}. Therefore, they are incapable to give a stringent criteria for distinguishing the inflation and big bounce at the early stage of the cosmological evolution. Fortunately, the dark matter particle makes an exemption for this limitation thank to its small cross section and heavy mass.  

The idea of using dark matter mass and its cross section as a smoking gun signal of the existence of the bouncing universe was firstly proposed in~\cite{Li:2014era}, in which the {\it thermally production \& strongly freezing-out} case and the {\it non-thermally production \& weakly freezing-out} case for the evolutions of dark matter in a bouncing universe are discussed.

In this paper, we consider the dark matter with a temperature-independent thermally averaged cross-section to study the evolution of its relic abundance through a generic bounce. We find a new avenue in which the dark matter is produced {\it thermally} and freeze-out {\it weakly} in the bouncing universe. It predicts a novel characteristic relation for the relic density and cross-section $\Omega_\chi\propto \langle \sigma v\rangle^0$ . {\it i.e.} its relic density is independent of its cross-section--in contrast to the well-known relation, $ \Omega_\chi\propto \langle \sigma v\rangle^{-1}$ for WIMP model in standard cosmology. Then we show this result open up a new possibility to satisfy the currently observed relic abundance, and would serve as a important signature for the big bounce universe scenario.

In the remainder of this paper, we also have investigated the case in which dark matter is produced {\it non-thermally } and freeze out {\it weakly}. We find its relic density is not only proportional to $\langle \sigma v\rangle$, but also proportional to the ratio of the bounce temperature and dark matter mass, $x_b^{-1}$. Again, it would also serve as a  signature for the big bounce universe scenario.

\section{Production of Dark Matter}

\begin{figure}[htp!]
\centering
\includegraphics[width=0.48\textwidth]{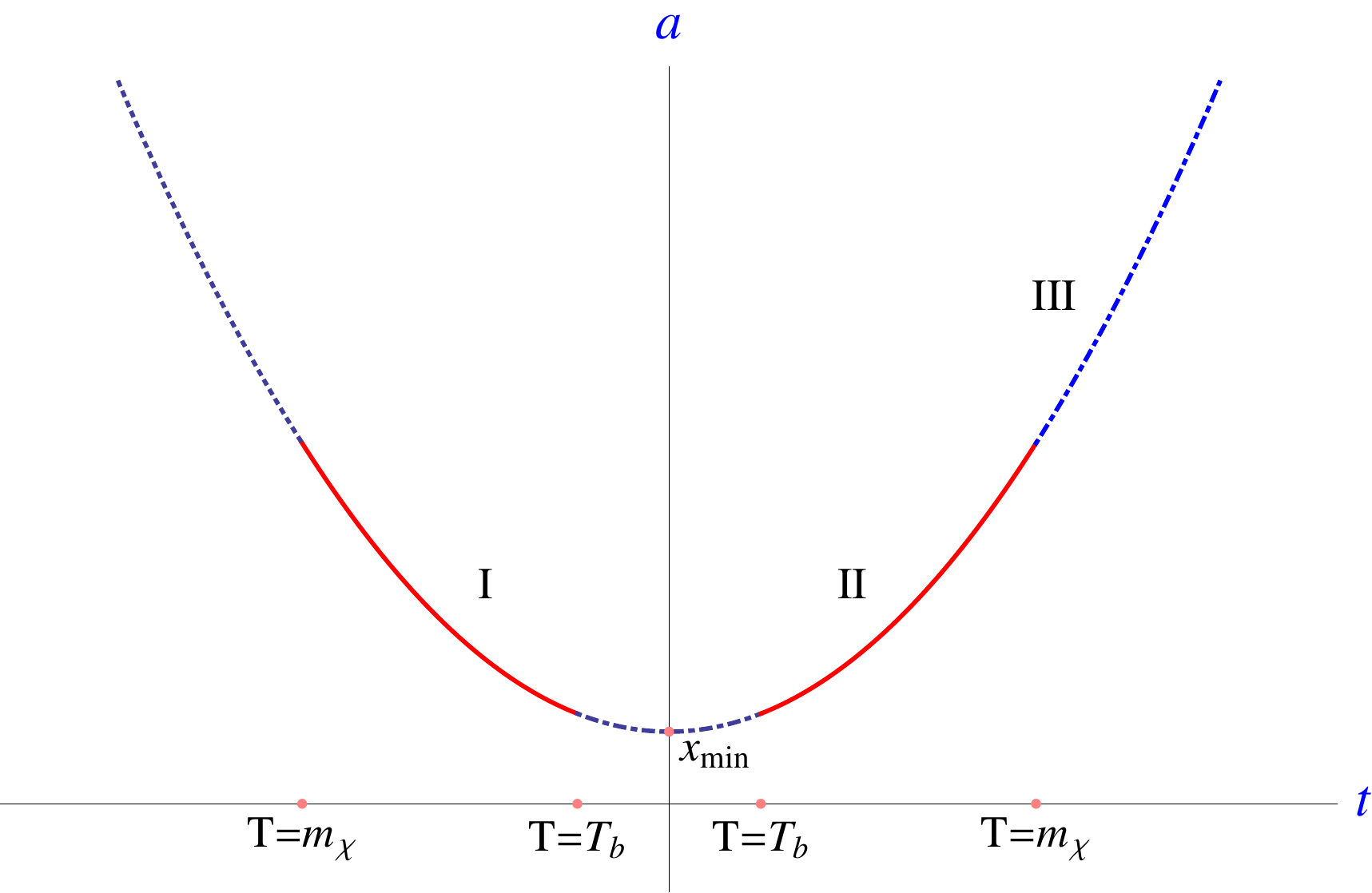}
\caption{The breakdown of the Big bounce period into a pre-bounce contraction (phase I), a post-bounce expansion (phase II), and  the  freeze-out  of the dark matter particles (phase III).
}
\label{fig:cosback.pdf}
\end{figure}

To facilitate a model independent analysis of the dark matter production in a generic bouncing universe scenario, we divide the bounce schematically into three stages~\cite{Li:2014era, Cheung:2014nxi} as shown in~ Figure. \ref{fig:cosback.pdf} : 
\begin{itemize}
\item{Phases I: {\it the pre-bounce contraction}, in which $H<0$ and $m_\chi< T < T_b$~;}
\item{Phases II: {\it the post-bounce expansion}, in which $H>0$ and $m_\chi< T < T_b$~;} 
\item{Phases III: {\it the freeze-out phase},  in which  $H>0$ and  $m_\chi \, >\,  T$~;}
\end{itemize}
and take a temperature-independent thermally averaged cross section $\langle \sigma v\rangle \propto T^0$, where $m_\chi$   is  the mass of dark matter, $\chi$, and $H$ the Hubble parameters taking positive value in expansion and negative value in contraction. $T$ and $T_b$ are the temperatures  of the cosmological background and of the bounce point, respectively. The bounce point, connecting Phases I and II with $T\sim T_b$,  is highly model-dependent.  The detailed modeling is sub-leading effect to our analysis of dark matter production as long as its time scale is short, and the bounce is assumed to be smooth and entropy conserved. 

Given that the entropy of universe is conserved around the bounce point~\cite{Cai:2011ci}, we, therefore, have a match condition for the relic abundance at the end of the pre-bounce contraction (denoted by $-$) and the initial abundance of the post-bounce expansion (denoted by $+$),
\begin{equation} \label{eq:match}
Y_- (x_b^-) = Y_+ (x_b^+)~, \quad Y\equiv\frac{n_\chi}{T^3}~, \quad x\equiv\frac{m_\chi}{T}~. 
\end{equation} 
where $n_\chi$ is the number density of dark matter particles. In the early stage before Phase I in the bouncing universe scenario, the temperature of background, $T\ll m_\chi$, is too low to produce dark matter particle efficiently, so the number density of dark matter particles can be set to zero at the onset of the pre-bounce contraction phase without of loss generality~\cite{Li:2014era}: 
\begin{equation} \label{eq:ini}
Y_-(T \sim m_\chi)= 0. 
\end{equation} 

The evolution of dark matter in a bouncing universe is governed by the Boltzmann equation,
\begin{equation} \label{eq:nsf}
\frac{d(n_\chi a^3)}{a^3dt}=\langle\sigma v\rangle\left[\left(n_\chi^{(0)}\right)^2-n_\chi^2\right]~, 
\end{equation} 
where $n_\chi^{(0)}$ is the equilibrium number density of dark matter, $a$  the scale factor of the cosmological background, and $\langle\sigma v\rangle$ the thermally averaged cross section. In accordance with the generic bounce universe scenario, we model the pre-bounce contraction and the post-bounce expansion phases to be radiation-dominated, $H\propto a^{-4}$. Then, in the pre-bounce contraction phase, Eq.\ref{eq:nsf} is simplified to be,
\begin{equation}
\frac{dY_-}{dx}=-f\langle \sigma v\rangle m_\chi x^{-2}(1-\pi^4Y_-^2)~, \label{eq:hn}
\end{equation}
where $f$ is constant during the radiation-dominated era, $f\equiv \frac{m_\chi^2}{\pi^2} (|H|x^2)^{-1}=6.01\times 10^{26}~eV$, as constrained by observations. Consequently, in the post-bounce expansion phase, Eq.\ref{eq:nsf} also simplifies
\begin{equation}
\frac{dY_+}{dx}=f\langle \sigma v\rangle m_\chi x^{-2}(1-\pi^4Y_+^2)~, \label{eq:hp}
\end{equation}
which differs Eq.\ref{eq:hn} by an overall sign $\pm$ due to the signs of Hubble constant in either expansion or contraction.  

\begin{figure}[htp!]
\centering
\includegraphics[width=0.48\textwidth]{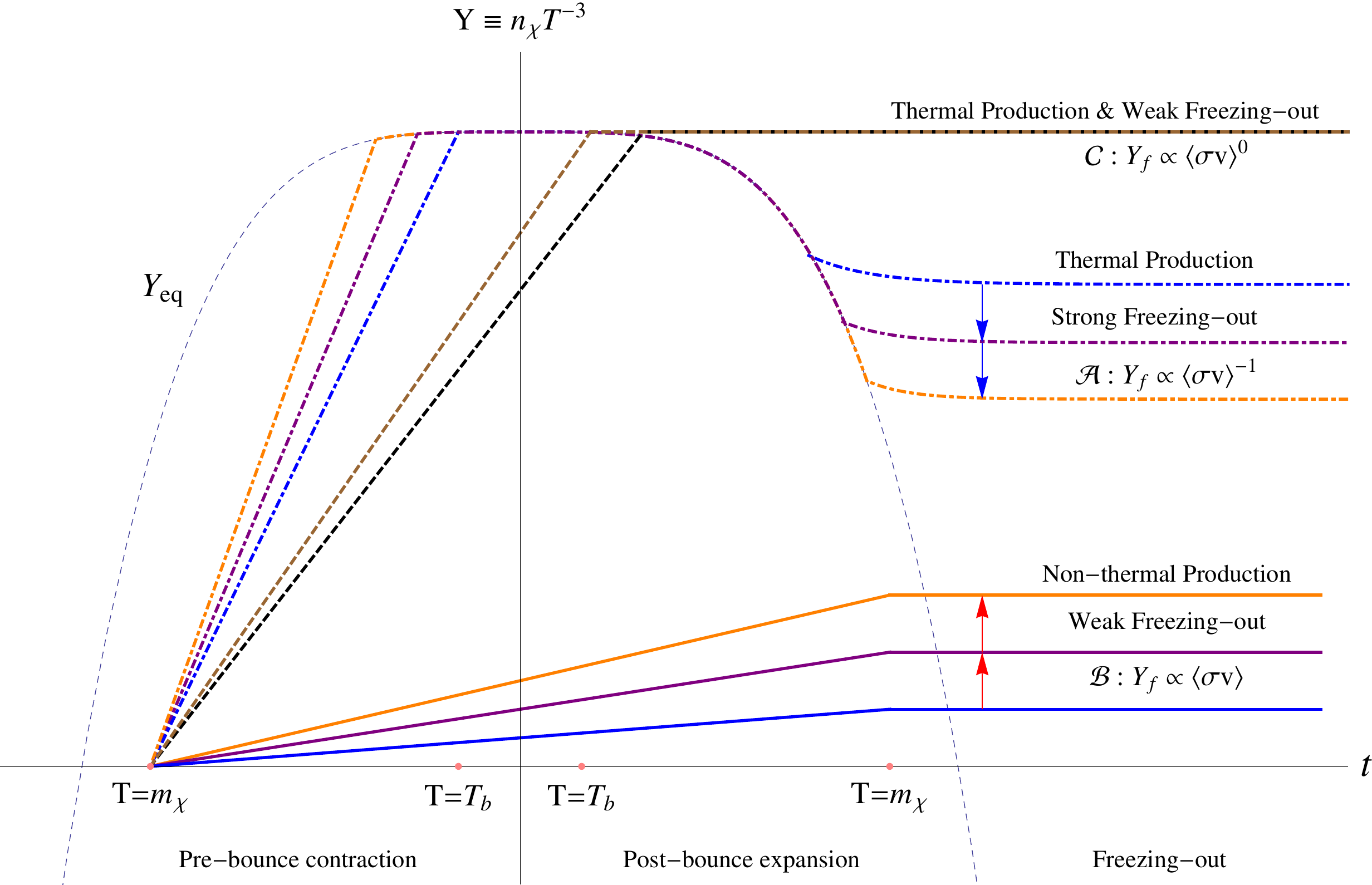}
\caption{A schematic plot of the time evolution of dark matter in a generic bounce universe scenario. Three outcomes producing dark matter for satisfying current observations are illustrated.
}
\label{fig:TriRelicEvolution.pdf}
\end{figure}

Solving Eq.\ref{eq:hn} and Eq.\ref{eq:hp} with the initial condition Eq.\ref{eq:ini} and Eq.\ref{eq:match} directly, we obtain the analytic solution of the dark matter abundance until the ending of the post-bounce expansion phase,
\begin{equation}
Y_+=\frac{1-e^{2\pi^2f\langle \sigma v\rangle m_\chi \left(\frac{1}{x}+\frac{x_b-2}{x_b}\right)}}{\left(1+e^{2\pi^2f\langle \sigma v\rangle m_\chi \left(\frac{1}{x}+\frac{x_b-2}{x_b}\right)}\right)\pi^2}~. \label{eq:yp}
\end{equation}
At the end of dark matter production, $T\sim m_\chi$, this complete solution can be categorized in two limits, {\it Thermal Production} and {\it Non-thermal Production} : 
\begin{equation}  \label{eq:Yplus2}
Y_+|_{x=1}=
\left\{  
  \begin{array} {lr}
 {\displaystyle \pi^{-2}, \qquad\qquad\qquad  4\pi^2f\langle\sigma v\rangle m_\chi x^{-1}_b\gg 1} 
 \\ 
  {\displaystyle 2f\langle\sigma v\rangle m_\chi x^{-1}_b~, \quad~ 4\pi^2f\langle\sigma v\rangle m_\chi x^{-1}_b\ll 1}   
  \\
\end{array}     
\right. .
\end{equation}
In {\it thermal production} case, the dark matter is produced swiftly to be in fully  thermal equilibrium with primordial plasma due to its large cross-section,  $4\pi^2f\langle\sigma v\rangle m_\chi x^{-1}_b\gg 1$. Then its abundance tracks the  equilibrium values, $\pi^{-2}$, until freezing out, as depicted in~ Figure. \ref{fig:TriRelicEvolution.pdf}.

And in the case of {\it non-thermal production}, the cross-section of dark matter is much smaller, $4\pi^2f\langle\sigma v\rangle m_\chi x^{-1}_b\ll 1$, so that the production of dark matter is insufficient to reach thermal equilibrium. Its abundance is proportional to $\langle\sigma v\rangle $, and the information of the cosmological evolution of the bouncing universe, the factor $2f x^{-1}_b~$, is carried on its outcome. Therefore, if such information can survive and be extracted after the freeze-out, it would become a signature of the bounce universe scenario.

\section{Freeze-out}
As universe cools sufficiently to be $T< m_\chi$ in the radiation dominated expansion,  dark matter undergoes a thermal decoupling and then freeze-out. In this freeze-out phase, Eq.\ref{eq:nsf} is simplified to be
\begin{equation}  \label{eq:dYdx}
\frac{d Y}{d x}=f\langle\sigma v\rangle m_\chi\left(\frac{\pi}{8}xe^{-2x}-\pi^4\frac{Y^2}{x^2}\right)~,
\end{equation}  
where the first term on the right hand side of Eq.\ref{eq:dYdx} is exponentially subdominant and is discarded for $x>1$. 
Integrating~Eq.\ref{eq:dYdx} from $x=1$ to $x\rightarrow\infty$, we obtain the relic abundance of dark matter after freeze-out,
\begin{equation}  \label{eq:dYdxsol}
Y_f\equiv Y|_{x\rightarrow\infty}=\frac{1}{\pi^4f\langle\sigma v\rangle m_\chi +(Y_+|_{x=1})^{-1}}~.
\end{equation}
Again, according to Eq.\ref{eq:dYdxsol} the relic abundance of dark matter indicates two distinctive patterns for the freeze-out process, {\it Strong} and {\it Weak} freeze-out (also see ~\cite{Li:2014era}).

{\it Strong freeze-out:} If the initial abundance of dark matter at beginning of the freeze-out process is very large,  $Y_+|_{x=1}\gg (\pi^4f\langle\sigma v\rangle m_\chi )^{-1}$, after freeze-out the relic abundance of dark matter $Y_f$ becomes independent of the initial abundance $Y_+|_{x=1}$ and inverse to the cross-section,
\begin{equation}
Y_f=\frac{1}{\pi^4f\langle\sigma v\rangle m_\chi}= \frac{0.17\times 10^{-28} eV^{-1}}{\langle\sigma v\rangle m_\chi}~.
\end{equation}
So the information of the early universe evolution carried in $Y_+|_{x=1}$ is almost washed out after such {\it strong freeze-out} process, and the relic abundance is significantly less than the initial abundance,  as it is also  the WIMP miracle and WIMP-less miracle cases in Standard Cosmology where the {\it strong freeze-out} condition is always assumed~\cite{Scherrer:1985zt, Feng:2008ya, Kolb:1990vq,Gondolo:1990dk, Dodelson:2003ft}. Therefore, the dark matter undergoes {\it strong freeze-out} would not provide any unique signature for the cosmological evolution of neither big bounce nor big bang. 

Furthermore, according to~Eq.\ref{eq:Yplus2}, the maximal initial abundance of dark matter produced at onset of the freeze-out process is $\pi^{-2}$, $Y_+|_{x=1}\le\pi^{-2}$. So the {\it strong freeze-out } condition, $Y_+|_{x=1}\gg (\pi^4f\langle\sigma v\rangle m_\chi )^{-1}$ , can be simplified to be $\langle\sigma v\rangle m_\chi\gg \pi^{-2}f^{-1}=1.68\times 10^{-28} eV^{-1}$.

\paragraph{Weak freeze-out} On the other hand, if, however,  the cross section and mass of dark matter is small, $\langle\sigma v\rangle m_\chi\ll 1.68\times 10^{-28} eV^{-1}$, the relic abundance of dark matter after freeze-out is just the initial abundance at onset of the freeze-out process,  
\begin{equation}
Y_f= Y_+|_{x=1}~.
\end{equation}
Therefore, the information of the cosmological evolution of the bouncing universe encode in $Y_+|_{x=1}$ is preserved in the relic abundance $Y_f$ after such {\it weak freeze out} process, and no significant decrease for the abundance of dark matter.   

\section{Relic Abundance}
According to above discussion, there are two different production routes and two distinct freeze-out processes, which give four possible combined avenues for the evolution of dark matter in bounce universe. But only three avenues,  {\it thermal} \& {\it strong}, {\it non-thermal} \& {\it weak} and {\it thermal} \& {\it weak} as illustrated in~ Figure.\ref{fig:TriRelicEvolution.pdf}, are viable,  because if dark matter is produced non-thermally, its abundance is too less to trigger the {\it strong freeze-out} process. 

These three viable avenues are summarized in Table~\ref{tab:dmfreezeout}. And with the relation $\Omega_\chi\propto Y_f$, they indicate three distinctive relations between the cross section $\langle \sigma v \rangle$ and relic density of dark matter, $\Omega_\chi$, 
\begin{itemize}
\item $\Omega_\chi\propto \langle \sigma v \rangle^{-1}$~, in which the dark matter produced in  {\it thermal production} freezes out  {\it strongly}, marked branch~$\mathcal{A}$~; 
\item $\Omega_\chi\propto \langle \sigma v \rangle$~, in which the dark matter produced in  {\it non-thermal production} undergoes the {\it weak} freeze-out, marked brach~$\mathcal{B}$~\footnote{A similar relation but within reheating framework has been analyzed in~\cite{Chung:1998rq,Chung:1998ua,Drees:2006vh}};
\item $\Omega_\chi\propto \langle \sigma v \rangle^0$~, in which the dark matter produced in  {\it thermal production} freezes out  {\it weakly}, marked branch~$\mathcal{C}$~. 
\end{itemize}

\begin{table*}[ht!]  
\vspace{-0.5cm}
\caption{\label{tab:dmfreezeout}Relic abundance of dark matter after freeze-out}
\begin{center}
\begin{tabular}{|c|c|c|}
\hline
& Thermal Production  & Non-thermal Production \\
& $4\pi^2f\langle\sigma v\rangle m_\chi x^{-1}_b\gg 1$ & $4\pi^2f\langle\sigma v\rangle m_\chi x^{-1}_b\ll 1$\\
\hline
Strong Freeze-out & ${\mathcal{A}}$~:~ $Y_f=\frac{0.17\times 10^{-28} eV^{-1}}{\langle\sigma v\rangle m_\chi}$& --- \\
$\langle\sigma v\rangle m_\chi\gg 1.68\times 10^{-28} eV^{-1}$& $m_\chi>216~eV$&\\
\hline
Weak Freeze-out & ${\mathcal{C}}$~:~ $Y_f=\pi^{-2}$ &${\mathcal{B}}$~:~ $Y_f=1.2\langle\sigma v\rangle m_\chi x_b^{-1}\times10^{27} eV$\\
$\langle\sigma v\rangle m_\chi\ll 1.68\times 10^{-28} eV^{-1}$& $\frac{x_b}{4\pi^2f}\gg\langle\sigma v\rangle m_\chi\ll \frac{1}{\pi^2f}$&$\langle\sigma v\rangle m_\chi\ll 0.42 x_b\times 10^{-28} eV^{-1}$\\
\hline
\end{tabular}
\end{center}
\vspace{-0.5cm}
\end{table*}

We notice that the branch~$\mathcal{C}$, in which dark matter is produced {\it thermally} and take a {\it weakly} freezing-out process, gives a novel relic density and cross-section relation of dark matter: $\Omega_\chi\propto \langle \sigma v \rangle^0$. It may imply a new type of dark matter candidate which have a nearly fixed mass and varying cross-section -- in contrast to the WIMP candidate which have a nearly fixed cross-section and varying mass. 

The branch~$\mathcal{B}$ is a typical avenue for dark matter evolution in big bounce, called as {\it big bounce genesis} in ~\cite{Li:2014era, Cheung:2014nxi}. In this branch, the abundance of dark matter encode the information of cosmological evolution of big bounce during the {\it non-thermal production}, and preserving this information after {\it weak freeze-out}. Precisely, the relic density is not only proportional to $\langle \sigma v \rangle$, but also proportional to $x_b^{-1}$, the ratio of the bounce temperature and dark matter mass. So its prediction, $\Omega_\chi\propto \langle \sigma v \rangle x_B^{-1}$, also can be viewed as a important signature of big bounce.     

And the branch~$\mathcal{A}$, $\Omega_\chi\propto \langle \sigma v \rangle^{-1}$, is indistinguishable with prediction of standard cosmology~\cite{Kolb:1990vq, Dodelson:2003ft} since after {\it strong freeze-out} all information of the early universe is washed out as discussed above.

\section{The observational constraints}

\begin{figure}[htp!]
\centering
\includegraphics[width=0.48\textwidth]{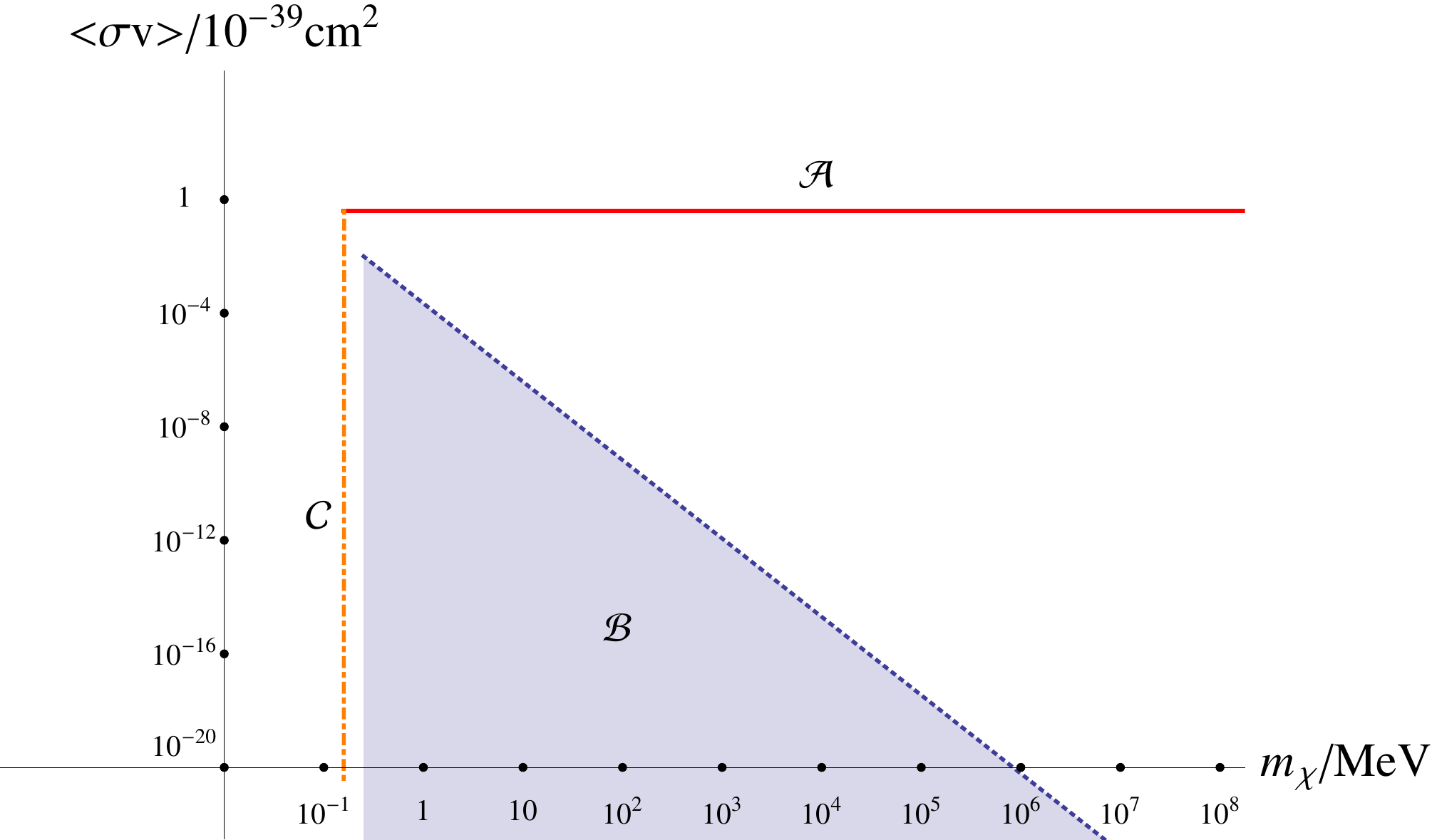}
\caption{Cosmological constraints on ${\langle\sigma v\rangle}$ and  $m_\chi$ of dark matter in the bounce universe scenario. The orange dot-dashed curve and  the shaded region is predicted by branch $\mathcal{C}$ and $\mathcal{B}$ respectively.}
\label{fig:Tricbp.pdf}
\end{figure}

To get a precise observational constraints, we imposing the currently observed value of $\Omega_\chi$, 
\begin{equation}
\Omega_\chi=1.18\times 10^{-2}eV\times m_\chi Y_f  =0.26~,
\end{equation}
into previous results, and obtain the cosmological constraints on ${\langle\sigma v\rangle}$ and  $m_\chi$ for branch~$\mathcal{C}$, ~$\mathcal{B}$, ~$\mathcal{A}$, as listed in Table~\ref{table:results} and plotted in Figure. \ref{fig:Tricbp.pdf}.

\begin{table*}[htdp]  
\vspace{-0.5cm}
\caption{\label{table:results} Cosmological constraints on ${\langle\sigma v\rangle}$ and 
 $m_\chi$ in the Bounce Universe Scenario.}
 \vspace{-0.0cm}
\begin{center}
\begin{tabular}{|c|c|}
\hline 
   &    $\Omega_\chi=0.26$ at present \\ 
\hline
Branch $\mathcal{A}$~:~ &   
${\langle\sigma v\rangle}=0.31\times 10^{-39}cm^2$~~,~~~  $m_\chi\gg 216~eV$\\ 
\hline
Branch ${\mathcal{B}}$~:~ &  
${\langle\sigma v\rangle}=1.82\times 10^{-26}m_\chi^{-2}x_b $~,~  $m_\chi\gg 432~eV$\\
\hline
Branch ${\mathcal{C}}$~:~~& $m_\chi=216~eV$~,~$ 0.42x_b\times10^{-28}eV^{-1}\ll \langle \sigma v\rangle\ll 1.68\times 10^{-28}eV^{-1}$\\
\hline
\end{tabular}
\end{center}
\vspace{-0.5cm}
\end{table*}

The branch $\mathcal{C}$ predicts a new falsifiable signature of the bounce universe scenario for satisfying the current observations,
\begin{eqnarray}  \label{eq:co}
\nonumber && m_\chi=216~eV~,\\
&& 0.42x_b\times10^{-28}eV^{-1}\ll \langle \sigma v\rangle\ll 1.68\times 10^{-28}eV^{-1},~~~
\end{eqnarray}
which is plotted as the orange dot-dashed curve in Figure.\ref{fig:Tricbp.pdf}. 

And  the branch $\mathcal{B}$ predicts another concrete relation
\begin{equation}\label{eq:bo}
{\langle\sigma v\rangle}=1.82\times 10^{-26}m_\chi^{-2}x_b, \quad m_\chi\ge 432~eV~, 
\end{equation}
for satisfying the current observations of $\Omega_\chi$, and it is also a falsifiable signature of the bounce universe scenario. 

Turn our attention to the experimental detection of dark matter particle. With $x_b\le1$, the predictions of branch $\mathcal{C}$ and $\mathcal{B}$, Eq.\ref{eq:co} and Eq.\ref{eq:bo}, become 
\begin{equation}\label{eq:ct}
 m_\chi=216~eV~, \langle \sigma v\rangle\ll 1.68\times 10^{-28}eV^{-1},
\end{equation}
and 
\begin{equation}\label{eq:bt}
\langle \sigma v\rangle m_\chi^2 < 1.82\times 10^{-26}, \quad m_\chi\ge 432~eV~,
\end{equation}
respectively in the $m_\chi$ and $\langle \sigma v\rangle$ parameter space. 

If in future experimentally measured $m_\chi$ and $\langle \sigma v\rangle$ satisfy either Eq.\ref{eq:ct} or Eq.\ref{eq:bt}, the orange dot-dashed curve or the shaded region in~ Figure. \ref{fig:Tricbp.pdf} , it strongly indicates that our universe went through a Big Bounce--instead of the inflationary scenario as postulated in Standard Cosmology--at the early stage of the cosmological evolution.

\section{Summary}
In this paper, we inverstigate the production and freeze-out of dark matter with a temperature-independent thermally averaged cross-section  in a generic bouncing universe framework. 

We report a new avenue of the evolution of dark matter in big bounce, which the dark matter is produced {\it thermally} and freeze out {\it weakly}. This avenue predicts a novel characteristic relation between the relic density and cross-section of dark matter $\Omega\propto \langle\sigma v\rangle^0$ --in contrast to the well-known relation, $ \Omega_\chi\propto \langle \sigma v\rangle^{-1}$ for WIMP model in standard cosmology. By imposing the currently observed value of $\Omega_\chi$, the relation of $\langle\sigma v\rangle$ and $m_\chi$  satisfying the current observations is obtained, $m_\chi=216~eV~, \langle \sigma v\rangle\ll 1.68\times 10^{-28}eV^{-1}$, which serves as a falsifiable signature of the bounce universe scenario and opens up a new possibility of experimentlly testing the bounce universe scenario using dark matter detection.

We also discuss the case in which dark matter is produced {\it non-thermally } and freeze out {\it weakly}. Its relic density is proportional to $\langle \sigma v\rangle x_b^{-1}$, and the cross-section and mass relation is ${\langle\sigma v\rangle}=1.82\times 10^{-26}m_\chi^{-2}x_b, ~ m_\chi\ge 432~eV$, which also serves as a falsifiable signature for the big bounce universe scenario with the property of dark matter particle.

To summarize, if the value of $m_\chi$ and $\langle \sigma v\rangle$ determined in near future dark matter detection experiments~\cite{Aprile:2012nq, Akerib:2013tjd, Aalseth:2010vx, Bernabei:2010mq, Adriani:2013uda, Cheung:2014pea, Battiston:2014pqa, Aguilar:2014mma,Xiao:2014xyn} satisfy either relations for this two avenues,  then it strongly indicates that bounce universe scenario are favorable.

\section{acknowledgments}
We would like to thank Yeuk-Kwan E. Cheung who launches this research project. Useful discussions with Jin U Kang,  Kai-Xuan Ni and Konstantin Savvidy are gratefully acknowledged. This research project has been supported in parts by the 
Jiangsu Ministry of Science and Technology 
under contract~BK20131264, by the 
Swedish Research Links programme of the Swedish Research Council (Vetenskapsradets generella villkor) under contract~348-2008-6049. We also acknowledge 
985 Grants from the Ministry of Education, and the 
Priority Academic Program Development for Jiangsu Higher Education Institutions (PAPD).

\section{Appendix: Cosmological Constraints on the Energy Scale of Big Bounce}
\begin{figure}[htp!]
\centering
\includegraphics[width=0.48\textwidth]{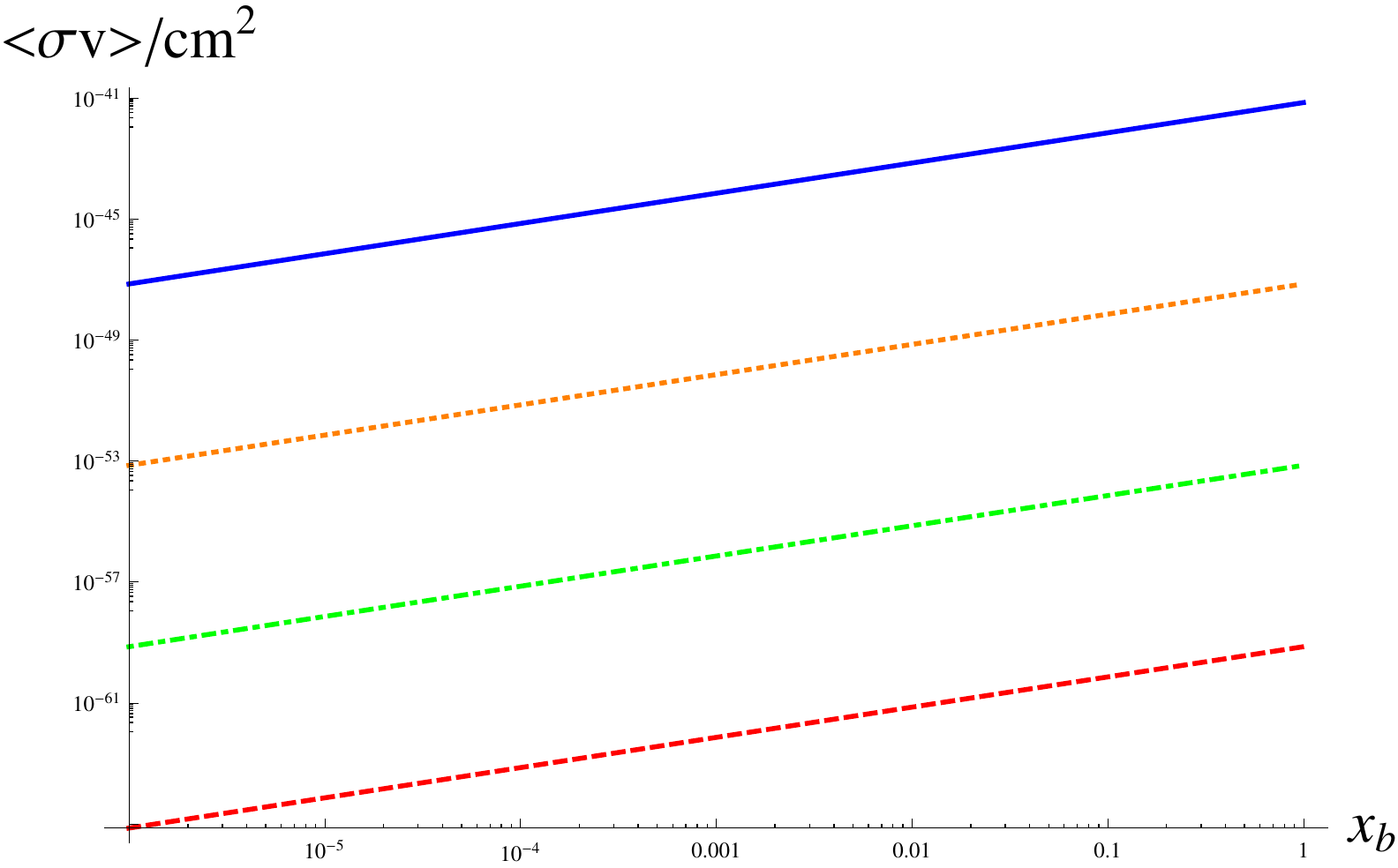}
\caption{Cosmological constraints on ${\langle\sigma v\rangle}$ and  $x_b$ for various fixed mass of dark matter particle--- 1~KeV(the blue solid line), 1~MeV(the orange dashed line), 1~GeV(the green dot-dashed line) and 1~TeV(the red dashed line) ---in the Bounce Universe Scenario.
}
\label{fig:xb.pdf}
\end{figure}

In Table \ref{table:results}, the cosmological constraints on the energy scale of big bounce, the cross section and mass of the dark matter for satisfying the current observation, $\Omega=0.26$, are obtained. In this appendix, we plot the constraints on the energy scale of big bounce and the cross section of dark matter for various fixed $m_\chi$ in accordance with the predication of Branch $\mathcal{B}$ in~ Figure. \ref{fig:xb.pdf}.

\bibliography{BBGref}

\end{document}